\documentclass[a4paper,12pt]{article}
\usepackage{amsfonts}
\usepackage{latexsym}
\usepackage{amsmath}
\usepackage{amssymb}
\usepackage{amssymb}
\usepackage{slashed}
\usepackage{upgreek}
\usepackage{mathrsfs}
\usepackage{bbm}
\usepackage{relsize}
\usepackage{graphicx}
\usepackage{pstricks}

\newcommand{\newsection}{    % Numeration of eqs. is automatic
\setcounter{equation}{0}\section}
\def\appendix#1{\addtocounter{section}{1}\setcounter{equation}{0}
\renewcommand{\thesection}{\Alph{section}}
\section*{Appendix \thesection\protect\indent \parbox[t]{11.15cm}{#1}}
\addcontentsline{toc}{section}{Appendix \thesection\ \ \ #1}}

\font\mybb=msbm10 at 11pt

\def\bb#1{\hbox{\mybb#1}}

\def\bZ {\bb{Z}}
\def\bR {\bb{R}}

\def\bC {\bb{C}}

\def\a {\alpha}
\def\b {\beta}
\def\g {\gamma}
\def\d {\delta}

\newcommand{\bea}{\begin{eqnarray}}
\newcommand{\eea}{\end{eqnarray}}

\begin{document}
\begin{titlepage}
\begin{center}
%\today
\vspace*{-1.0cm}
%\hfill hep-th/yymmnnn \\
%\hfill UB-ECM-PF-06-43 \\

{\Large \bf Patching DFT, T-duality and Gerbes} \\[.2cm]

\vskip 2cm
 P.S. Howe and G. Papadopoulos
\\
\vskip .6cm

\begin{small}
\textit{  Department of Mathematics,
\\
King's College London,
\\
Strand, London WC2R 2LS, UK.\\
E-mail: paul.howe@kcl.ac.uk ; george.papadopoulos@kcl.ac.uk}
\end{small}\\*[.6cm]

\end{center}

%\today

\vskip 3.5 cm
\begin{abstract}
We clarify the role of the dual coordinates as described from the perspectives of the Buscher T-duality rules
and Double Field Theory.  We show that the T-duality angular dual coordinates cannot be identified with Double Field Theory dual coordinates in any
of the proposals that have been made in the literature for patching the doubled spaces.
In particular, we show with  explicit examples that the T-duality  angular dual coordinates  can  have non-trivial transition functions over a spacetime  and that their identification
with the Double Field Theory dual coordinates is
in conflict with proposals in which the latter remain inert under the patching of the B-field.  We then  demonstrate that   the Double Field Theory coordinates can be identified
with some C-space coordinates  and that  the T-dual spaces of a spacetime are  subspaces of the gerbe in  C-space. The construction  provides a description of both the local $O(d,d)$ symmetry and the T-dual spaces
of spacetime.

\end{abstract}

\end{titlepage}

%%%%%%%%%%%%%%%%%%%%%%%%%%%%%%%%%%%%%%%%%%%%%%%%%%%%%%%%%%%%%%%%%%%%%%%%%%

%%%%%%%%%%%%%%%%%%%%%%%%%%%%%%%%%%%%%%%%%%%%%%%%%%%%%%%%%%%%%%%%%%%%%%%%%%
\section{Introduction}

One of the requirements for the consistent formulation of double and exceptional field theories  is a description
of the patching conditions of doubled and exceptional spaces that underpin these theories. Let us for simplicity focus
on doubled spaces as many more results for these are known. Doubled spaces arise by adding to the spacetime coordinates
$x$ a set of dual coordinates $\tilde x$. In double field theory (DFT), the new coordinates are as many as those of the spacetime.

The question that arises is how these new coordinates patch. There are two main  approaches in the literature for this.
In the first approach, it is proposed that the dual coordinates $\tilde x$
patching transformations depend on the transition functions of the $B$-field. There are various suggestions for such dependence.  Two such suggestions can be found in
\cite{lust} and \cite{rey}.

 Another  approach, advocated in \cite{hull}, asserts  that the patching conditions of the dual coordinates $\tilde x$ can be arranged such that they do not depend
on the transition functions of the $B$-field. In such a case, the doubled space of any string background spacetime $M$ is either a product space $M\times Q$, where $Q$ can be chosen as  $\bR^n$ or $T^n$, or the cotangent bundle $T^*M$.
The dual coordinates in this case  become forgetful, in the sense that they are inert under $B$-field gauge transformations.

 In addition, a recent proposal for DFT for some coset spaces was made in \cite{Hassler:2016srl}, following on from an analysis of DFT for Wess-Zumino-Witten models presented in \cite{Blumenhagen:2014gva,Blumenhagen:2015zma}.

One of the difficulties in deciding the way that the dual coordinates should patch  is the  uncertainty of which criteria one should apply.
A selection of such criteria is as follows:

\begin{itemize}

\item The doubled spaces patch in such a way  that is consistent with the dual spaces obtained via the Buscher T-duality rules.

\item The patching of double spaces is such that it requires for consistency the Dirac quantisation property of the 3-form flux.

\item The doubled spaces satisfy the topological geometrisation condition.

\item Doubled spaces can be constructed for all  backgrounds with 3-form flux.

\item Generalised geometry emerges naturally   on doubled spaces.

\end{itemize}

The first criterion is perhaps the most conservative one. Whatever the patching of doubled spaces is, it should reproduce both locally and globally the
results that arise after applying  the Buscher T-duality rules.  After all these produce the only explicit examples we know.
Locally this is indeed the case  through the use of $O(d,d)$ duality transformations \cite{duff, tseytlin, siegel} on the fields. However, we shall see that globally the patching conditions of the doubled
spaces do not reproduce the results obtained from Buscher rules.

Moreover, it is worth mentioning that DFT has raised the expectations of what can be described.  As the transformations
 of  DFT  make no mention of isometries that are instrumental  in the Buscher rules, there is
some expectation that the doubled spaces can be used to describe a dual space which arises after dualising all   spacetime directions. Another aspect of the dualisation
of the whole spacetime is the idea of geometrisation, i.e. the notion that the theory which includes the spacetime metric and the 3-form field strength
can be described in terms of metric data only.  This is analogous to Kaluza-Klein theory which provides a geometrisation for a 2-form field strength.

The second criterion is an extrapolation of a similar result that arises in Kaluza-Klein theory. The construction of the Kaluza-Klein space
is achieved after restricting the 2-form field strength to represent the first Chern class of a line bundle. In turn the flux of the 2-form
is required to obey the Dirac quantisation condition.

The third criterion is also posed in analogy with the Kaluza-Klein theory.  It states that the pull-back of the 3-form field strength on
whatever a consistent description of doubled space is, or a generalisation of it, must represent the trivial cohomology class \cite{gp2}.  This has several
consequences such as, for example, that the dual coordinates must have a non-trivial topology and non-trivial transition functions over the spacetime.

The fourth criterion is a natural one from the point of view of DFT.  In all proposals made in the literature for the theory,
there is no restriction mentioned on the backgrounds.

The fifth criterion is introduced because in generalised geometry the T-duality group $O(d,d)$ arises naturally as the (sub)group of automorphisms
of a vector bundle.  So the expectation is that in a consistent formulation of the doubled space this bundle should arise naturally. In fact
it is expected to be related to, if not identified with,  its tangent bundle.

There are several proposals in the literature on how the doubled spaces might patch and some analysis of how they measure against the criteria mentioned above. In particular, the  patching of doubled spaces under the transformations proposed in \cite{lust} has been investigated in \cite{gp1} where it was shown
that consistency  on 4-overlaps requires that the 3-form field strength $H$ must be exact.   To resolve the patching issue, C-spaces, essentially local descriptions of gerbes,  have been proposed in \cite{gp2}. They exhibit consistent patching with a cohomologically non-trivial $H$ and locally contain the doubled spaces,  but generically  they have
more coordinates than doubled spaces. Indeed, in the case of non-trivial $H$-fields they do not have well-defined global dimensionalities.

More recently, two new proposals for patching doubled spaces  have been put forward \cite{rey}, \cite{hull}. In this paper, we shall consider these two proposals
and investigate them in the light of  the criteria mentioned above. First we shall clarify some aspects of the patching conditions proposed in \cite{rey} and  demonstrate that, up to an
allowed redefinition of the dual coordinates and choice of transition functions for $B$ at double overlaps,
the patching conditions of the dual coordinates do not depend on the transition functions of the $B$-field. As a result, for these choices, the dual coordinates of the doubled space remain inert
under patching which in turn implies that this proposal is related to that of \cite{hull}.

The proposal made in \cite{hull} states that the dual coordinates of a doubled space remain inert under patching and the transformations induced by the form part of a generalised vector acting infinitesimally with a
generalised Lie derivative on the fields are not
coordinate transformations but rather gauge transformations of the $B$-field. As a result the dual coordinates can be forgetful and the spacetime geometry is described by
a generalised geometry structure and a splitting of the generalised geometry bundle induced by the $B$ field interpreted as a gerbe connection.

 In the proposal of \cite{Hassler:2016srl} for DFT on group manifolds, the doubled space is a group manifold with the physical space embedded into it as a Lagrangian type of submanifold, after the strong section condition is imposed, while the T-dual space corresponds to a different embedding. In this case both the physical and dual coordinates are non-trivially patched.

In what follows we give a detailed analysis of the T-duality pair of  $S^3$ with $N$ units of $H$-charge and the lens space $L^3_N=S^3/\bZ_N$ with $1$ unit of $H$ charge. We show
that the dual circle   twists topologically non-trivially over the spacetime $L^3_N$ and  therefore that either DFT dual coordinates cannot be identified\footnote{As our results are topological,
 this rules out all continuous and even  homotopic identifications.} with the T-duality angular coordinates,
or that the doubled spaces patching proposed in \cite{rey} and \cite{hull} is not consistent globally with the T-duality rules.
We also generalise this to other T-dual pairs including an example of T-dual spaces constructed from the 3-torus with H-flux background.

 Note that a conflict between T-duality and the strong section condition in doubled spaces had been pointed out before from a different perspective in \cite{martin, martin2}.
There a resolution was proposed by allowing additional transformations which preserve the split signature metric on the doubled space but  do not satisfy
the strong section condition.

 We then go on to
 propose a scenario   based on C-spaces and the Hitchin-Chatterjee  definition  of a  gerbe  in which  both the local $O(d,d)$ symmetry and the Buscher T-dual spaces
can be consistently described.  We propose an identification of
the DFT  coordinate $\tilde x$ of \cite{hull}, which transforms as a 1-form, with a coordinate that arises in the C-space construction \cite{gp2}. We then demonstrate how the T-dual space $\tilde M$
of a spacetime $M$ with  $H$-flux and which is a circle fibration can be identified as a subspace of the total space of the gerbe associated to $H$ on $M$.   We also provide explicit examples
of this which include the description of the T-dual lens space $L^3_N=S^3/\bZ_N$  of  $S^3$ with $N$ units of $H$-charge as a subspace of the total space of a
  gerbe on $S^3$.   The latter can be described as the union of $S^3$ with a circle bundle with first Chern class $N$ over an open neighbourhood of the equatorial
$S^2$ of $S^3$. The $L^3_N$ subspace of the gerbe is the restriction of this circle bundle over the equatorial $S^2$ of $S^3$.
We also give a similar construction for a T-dual space associated  3-torus background  with H-flux.  As the angular coordinates that arise naturally in  the gerbe construction,
and which are required for the identification of the T-dual spaces of spacetime as subspaces of gerbes, are not included in doubled spaces and therefore not in  DFT, we conclude that, for the
  consistent description of a theory with manifest Buscher T-duality symmetry, additional coordinates are required  in addition to  those of doubled spaces.

 The  paper is organised as follows: in section 2, we give the necessary and sufficient conditions for the T-dual circle to (topologically) twist over a spacetime  in a manner consistent with the Buscher rules.  We also prove
that the dual circle of  the lens space $L^3_N$, viewed as a circle fibration over $S^2$, and that of $T^3$ with $H$-flux, topologically twist over the spacetime.  In section 3, we review the proposals for patching DFT that have appeared
in the literature
and in section 4 we investigate them from a patching point of view  concluding that they do not describe the topological twist of the dual circles.  In section 5, we explore the relation between
doubled spaces and C-spaces, explain how local $O(d,d)$ symmetry arises,  and  present a gerbe
construction for all spacetimes which are circle fibrations and have  some $H$-flux which allows for the identification of the T-dual space as a subspace of the gerbe.
We also present explicit examples based
on $S^3$ and $T^3$ with H-flux backgrounds. In section 6, we present our conclusions.

\newsection{T-duality Rules and Patching}

\subsection{T-duality rules}

To describe the Buscher T-duality rules one assumes that the spacetime $M$ admits an $S^1$ group action  generated by a vector field $X$ which leaves the common
sector fields, the metric $g$, 3-form field strength $H$ and dilaton $\Phi$, invariant. Adapting coordinates along $X={\partial\over \partial\theta}$, the metric and 2-form
gauge potential can be written as
\bea
ds^2=V^2 (d\theta+q_i dx^i)^2+  {g}_{ij} dx^i dx^j~,~~~B=(d\theta+q_i dx^i)\wedge p_j dx^j+{1\over2} b_{ij} dx^i\wedge dx^j~.
\label{tduala}
\eea
After performing a T-duality transformation, the dual metric,  2-form gauge potential and dilaton  read
\bea
d\tilde s^2&=&V^{-2} (d\tilde\theta+p_i dx^i)^2+  {g}_{ij} dx^i dx^j~,~~\tilde B=(d\tilde\theta+p_i dx^i)\wedge q_j dx^j+{1\over2} b_{ij} dx^i\wedge dx^j~,
\cr
e^{2\tilde \Phi}&=&e^{2 \Phi} V^{-2}~,
\label{tdualb}
\eea
where a new angular coordinate $\tilde \theta$ has now been introduced.  This is referred as the T-dual coordinate of $\theta$ and the associated circle as the dual circle, which we denote $\tilde S^1$. The coordinates
of $(x^i, \tilde \theta)$ are those of a new spacetime $\tilde M$ which, apart from having different geometry,  can also have different topology to that of $M$.
Furthermore $\tilde M$ again admits a $\tilde S^1$ action given by translations in $\tilde \theta$.
Another significant issue, which will be of central focus in what follows, is that the T-dual coordinate $\tilde \theta$ can have non-trivial
patching conditions over the original spacetime $M$ (or vice versa). These are given by some of the transition functions of the $B$-field.  As can be seen from (\ref{tduala}), $p_i$ will transform under a $B$-field transformation and this will induce a transformation of $\tilde\theta$ in (\ref{tdualb}) in order for the T-dual metric to remain invariant.

 The original spacetime $M$ together with its dual $\tilde M$ can be put together to construct an enhanced space.  To see this observe that the space of orbits of the $S^1$ action on $M$
and of the $\tilde S^1$ action on $\tilde M$ are the same, $M/S^1=\tilde M/\tilde S^1=Q$. To avoid complications with fixed points, let us assume from now on that
the action of $S^1$ on both spaces is free\footnote{Otherwise, one can use the slice theorem to remove the fixed points and repeat the same analysis on the remaining space.}.
In such a case, one can construct a torus bundle $P(Q,T^2)$ over $Q$. The torus bundles are classified by elements in $H^2(Q, \bZ)\oplus H^2(Q, \bZ)$ which
are the first Chern classes of $M$ and $\tilde M$ viewed as circle bundles over $Q$.  In \cite{hull2}, $P(Q,T^2)$ is referred to as the correspondence space.
In particular, the first Chern classes are represented by the 2-forms ${1\over 2\pi} dq$ and ${1\over 2\pi} dp$, respectively, with $p=p_i dx^i$ and similarly for $q$.

We therefore have the diagram

\begin{equation}
\begin{picture}(400,100)
%\put(202,180){(2,2)}
%\put(210,175){\vector(-1,-1){20}}\put(215,175){\vector(1,-1){20}}
%\put(170,145){(2,1)}\put(235,145){(1,2)}
%\put(170,135){\vector(-1,-1){20}}\put(180,135){\vector(1,-1){20}}\put(245,135){\vector(-1,-1){20}}\put(255,135){\vector(1,-1){20}}
%\put(135,100){(2,0)}
\put(210,100){$P$}
%\put(275,100){(0,2)}\put(280,95){\vector(-1,-1){20}}
%\put(145,95){\vector(1,-1){20}}
\put(185,85){$S^1$}\put(235,85){$\tilde S^1$}
\put(170,60){$\tilde M$}
\put(210,95){\vector(-1,-1){20}}
\put(215,95){\vector(1,-1){20}}\put(240,60){$M$}
\put(190,55){\vector(1,-1){20}}\put(235,55){\vector(-1,-1){20}}
\put(185,40){$\tilde S^1$}
\put(235,40){$ S^1$}
\put(210,20){$Q$}

\end{picture}
\label{4.13.b}
\end{equation}

We can also define two-forms $F$ and $\tilde F$ on $Q$ by integrating $\tilde H$ over $\tilde S^1$ and $H$ over $S^1$ respectively. Here, from the T-duality rules,
\bea
&&H=dB=-d\theta\wedge dp + h +d(q\wedge p)
\cr
&&\tilde H=d\tilde B=-d\tilde\theta\wedge dq + h +d(p\wedge q)\ ,
\label{2.4}
\eea
where $h=db$. So
\bea
F=-\frac{1}{4\pi^2} \int_{\tilde S^1} \tilde H\ ;\qquad \tilde F=-\frac{1}{4\pi^2}\int_{ S^1} H\ .
\label{2.5}
\eea

Equations (\ref{2.4}) and (\ref{2.5}), together with the fact that both $M$ and $\tilde M$ have the same quotient $Q$ as circle bundles, were specified as the required conditions for the two spaces to be T-dual in \cite{Bouwknegt:2003wp}.

\subsection{The T-dual circle topologically twists over the spacetime}

Although the T-dual coordinates $\tilde \theta$ have non-trivial transition functions over $Q$, it does not necessarily mean that they are
(topologically) twisted over the spacetime $M$. To settle this question, let us examine an example in detail.  This is the well-known
T-dual pair of $S^3$ with N-units of $H$ flux and the 3-dimensional lens space $L^3_N$ with 1-unit of $H$ charge. It is useful to note that $L^3_N$ is the space of orbits of $\bZ_N$ on $S^3$ where the generator $g=\exp {2\pi i/N}$ of $\bZ_N$ acts as $v_r\rightarrow  gv_r$, where $v_r$ are complex numbers such that
$v_1\bar v_1+v_2 \bar v_2=1$.

Both spaces $S^3$ and $L^3_N$ are circle fibrations over $S^2$,  $Q=S^2$. Moreover the first Chern class of these fibrations is $c_1(S^3)=u$ and  $c_1(L^3_N)= N u$, respectively,
where $u$ is the generator of $H^2(S^2, \bZ)$.  Furthermore the cohomology groups of $S^3$ and $L^3_N$ are
\bea
&&H^0(S^3, \bZ)=H^3(S^3, \bZ)=\bZ~,~~~H^1(S^3, \bZ)=H^2(S^3, \bZ)=0~,~~~
\cr
&&H^0(L^3_N, \bZ)=H^3(L^3_N, \bZ)=\bZ~,~~~H^1(L^3_N, \bZ)=0~,~~~H^2(L^3_N, \bZ)=\bZ_N~.
\eea
Next consider the $T^2$ fibration $P=P(T^2, S^2)$  with first Chern classes $c_1(P)=u$ and $c_1(P)=N u$.  In fact  $P=\big(S^1\times S^3\big)/\bZ_N$, where now the generator $g$ of $\bZ_N$ acts as $(a, v_r)\rightarrow (g a , g v_r)$ and $|a|=1, a\in \bC$. It turns out that the cohomology of $P$ can be computed and can be found that
\bea
H^0(P, \bZ)=H^1(P, \bZ)=H^3(S^3, \bZ)=H^4(P, \bZ)=\bZ~,~~H^2(P, \bZ)=0~.
\label{pcoh}
\eea
In particular observe that the middle cohomology of $P$ vanishes.

To continue observe that $P$ can be viewed as a circle fibration over either $S^3$ or $L^3_N$.  Consider first $P$ as a circle fibration over $S^3$.
This fibration is obtained  after considering the group action $[a, v_r]\rightarrow [a z, zv_r]$, where  $z\in S^1\subset \bC, |z|=1$ is the group element and $[a, v_r]$ denotes the orbit of $Z_N$
represented by $(a, v_r)$.
In fact
notice that $S^1/\bZ_N=S^1$ acts freely.
As $H^2(S^3,\bZ)=0$, all circle bundles
over $S^3$ are topologically trivial.  As a result $P$ is a topological product $S^1\times S^3$.  One therefore concludes that the dual coordinate $\tilde \theta$ {\it does not twist} over the
spacetime $S^3$.

However  the T-dual Lens space $L^3_N$ can also be considered as the spacetime, and so $S^3$ can be thought as its T-dual. Note that $H^2(L^3_N, \bZ)=\bZ_N$ and so $L^3_N$ admits topologically non-trivial circle bundles.  The fibration of $P$  over $L^3_N$ is constructed by considering  the circle action $[a, v_r]\rightarrow [a z, v_r]$.  If $P$ was
a trivial topological product $S^1\times L^3_N$, the K\"unneth formula for computing the cohomology of the topological product of two spaces would have implied that
 \bea
 H^2(P, \bZ)= H^2(L_N^3, H^0(S^1, \bZ))=H^2(L_N^3, \bZ)=\bZ_N~.
 \eea
 This is a {\it contradiction} as the second cohomology of $P$ vanishes (\ref{pcoh}).
Therefore $P$ is a topologically twisted product of $S^1$ and $L^3_N$.  As a result, the dual $\theta$ coordinate has {\it non-trivial patching conditions}  over the spacetime $L^3_N$.

Incidentally, observe that $P$ satisfies  a partial version of  the topological geometrisation condition of \cite{gp2}.
Both the $S^3$ backgrounds and its dual $L^3_N$ have non-trivial
$H$ fluxes. As a result, the T-duality operation does not geometrise all of the B-flux, so that one does not expect that the pull back of $H$ or $\tilde H$ on $P$ will represent the trivial class in $H^3(P, \bZ)$. Instead the topological geometrisation condition manifests itself as follows: pulling  back $H$ and $\tilde H$ onto $P$, one may have expected that these represent two independent
cohomology classes in $H^3(P, \bZ)$, but  this is not the case. $H^3(P, \bZ)$ has one generator and the linear combination
$N H-\tilde H$ represents the trivial class in $H^3(P, \bZ)$, where we have suppressed the pull-back operations.  This is because part of the information of the transitions
functions of $H$ and $\tilde H$ is stored in the patching conditions of $P$.

The example we have given  above can be generalised to include $T^n$ actions and thus T-duality in more than one direction. However, for the purpose
of this paper, the example we have investigated will suffice.

To conclude, {\it the Buscher T-duality rules allow for the possibility that the dual circle has a non-trivial topological twist over the spacetime},
 so that the dual angular coordinates can  have {\it non-trivial patching conditions} over the spacetime.  As we have seen, this situation does indeed arise in explicit examples.

\subsection{A patching approach to T-duality}

To give a bit more insight into the construction of circle bundle over a space  and its relation to the T-dual pairs, let us first describe how the third
cohomology group of the spacetime is constructed from the cohomology of $S^1$ and that of $Q$. \footnote{ In this subsection we allow $Q$ to have more than two dimensions in the general discussion.}
Assuming again that $S^1$ acts freely on the
spacetime $M$ and that $Q$ is simply connected, one can use the method of spectral sequences to determine $H^3(M,\bZ)$ from $H^1(S^1, \bZ)$,
$H^2(Q, \bZ)$ and $H^3(Q,\bZ)$.  The construction is rather intuitive. The elements of $H^3(M, \bZ)$ either are generated by $a u$, where $a$ is the generator of
in $H^1(S^1, \bZ)$ and $u$ are generators of $H^2(Q)$, or they are pulled-back from elements in $H^3(Q, \bZ)$ with the projection map.  This is precisely the case if $H^4(Q, \bZ)=0$.
If on the other hand $H^4(Q, \bZ)\not=0$, then only some of classes generated by  $a u$ may represent elements in $H^3(M, \bZ)$.  In either case, the 3-form
field strength $H$ in cohomology can be written as $[H]= a w+ v$, where $w\in H^2(Q, \bZ)$ and $v\in H^3(Q, \bZ)$, and where  the pull-back operation on $v$
has been suppressed.

It is clear from the T-duality rules stated in (\ref{tduala}) and (\ref{tdualb}) that the component of $H$ that take an active part in the
T-duality transformations is represented by $ a w$. Assuming that $w\in H^2(Q, \bZ)$, the dual space $\tilde M$ as a circle bundle  has first Chern class
$w$.
For  later applications, let us assume that $w$ is represented by a 2-form $\tilde F^2$.  The construction of $\tilde M$ can be made using a good cover
$\{U_\alpha\}_{\alpha\in I}$ on $Q$.  Then, using the Poincar\'e lemma on $U_\alpha$, and on  double and triple overlaps, $U_{\alpha\beta}=U_\alpha\cap U_\beta$  and
$U_{\alpha\beta\gamma}=U_\alpha\cap U_\beta\cap U_\gamma$ respectively, we find
\bea
\tilde F_\alpha^2= d C^1_\alpha~,~~~-C^1_\alpha+C^1_\beta= da^0_{\alpha\beta}~,~~~a^0_{\beta\gamma}-a^0_{\alpha\gamma}+a^0_{\alpha\beta}=n_{\alpha\beta\gamma}~,
\eea
where $C$ is the 1-form gauge potential, $a^0$ are the transition functions on double overlaps and $n$ are constants.  The latter lie in $2\pi \bZ$ as ${1\over 2\pi}\omega^2$
represents a class in $H^2(Q, \bZ)$.  Then $\tilde M$ is constructed by introducing an angular coordinate $\tilde \theta$ and after imposing the
patching conditions
\bea
\tilde\theta_\alpha-\tilde\theta_\beta-a^0_{\alpha\beta}=0~~ \mathrm{mod}~ 2\pi \bZ~.
\eea
These patching conditions are consistent on triple overlaps as $n_{\a\b\g}\in 2\pi \bZ$.

Making use of the above, we can state the criterion for whether the dual angular coordinate has non-trivial transition functions over the spacetime. Indeed, writing
$[H]= a w+ v$ and $[\tilde H]= a \tilde w+ \tilde v$,   we observe that  the dual angular coordinate $\tilde \theta$ has non-trivial transition functions over the spacetime iff $w$ represents a non-trivial class
in $H^2(M, \bZ)$, where the pull-back operation from $H^*(Q, \bZ)$ to $H^*(M, \bZ)$ has been suppressed.  Similarly, the angular  coordinate $\theta$ has non-trivial transition functions over the dual space $\tilde M$
iff $\tilde w$ represents a non-trivial class in $H^2(\tilde M, \bZ)$.  The classes $w$ and $\tilde w$ are represented by the forms $\tilde F$ and $F$ in (\ref{2.5}) respectively.

\subsection{T-duality on $T^3$ with flux}

We can use the results of the previous section to demonstrate that the T-dual angular coordinate of $T^3$ with flux also is twisted over the spacetime.
For this denote the angular coordinates of $T^3$ with $(\psi_1, \psi_2, \psi_3)$, $0\leq \psi_i<2\pi$, $i=1,2,3$. The metric and flux are given as
\bea
ds^2=(d\psi_1)^2 +(d\psi_2)^2+(d\psi_3)^2~,~~~H=-{N\over 4\pi^2} d\psi_1\wedge d\psi_2\wedge d\psi_3~,
\eea
where $N\in \bZ$.
If we choose as a T-duality direction $\psi_1$ and solve for the gauge potential as $B={N\over 4\pi^2} \psi_2 d\psi_1\wedge d\psi_3$, then
\bea
p={N\over 2\pi} \psi_2 d\psi_3~,
\eea
where the Killing vector field along the T-duality has been normalised as $2\pi \partial_{\psi_1}$. As it has been explained in the previous section,
the dual coordinate topologically twists over the spacetime iff the pull-back of $dp$ represents a non-trivial cohomology class. Indeed
\bea
dp={N\over 2\pi} d\psi_2\wedge d\psi_3~,
\eea
and its pull-back on $T^3$ is a non-trivial class as ${1\over 2\pi} d\psi_2\wedge d\psi_3$ represents one of the three generators of $H^2(T^3,\bZ)$.

\newsection{Double Field Theory Finite Transformations}

There has been extensive work in the literature to determine the allowed finite transformations of DFT. A concise description of all
possibilities and the sources can be found in \cite{hull}.  Here after imposing the strong section condition, we shall briefly summarise  the finite transformations proposed
as well as their induced action on the $B$-field.  This will suffice for the purpose of the analysis that follows
below.

First let us begin with the proposal of \cite{lust}.  In this proposal, the doubled space coordinates $(x^i, \tilde x_i)$ transform as
\bea
x'{}^i=x'{}^i(x^j)~,~~~~\tilde x'_i=\tilde x_i- v_i(x)~,
\label{dspacet}
\eea
and the induced transformation on the $B$ field is
\bea
B'_{ij}(x')&=& {\partial x^k\over \partial x'^i}\, {\partial x^l\over \partial x'^j} \, \Big(B_{kl}(x)+{1\over2}\big(
 {\partial v_{ l}\over\partial x^k} -
 {\partial v_{ k}\over\partial x^l} \big)\Big)
 \cr
 &&~~~+{1\over2} \Big( {\partial x^k\over \partial x'^i}\, {\partial v_{ j}\over\partial x^k}  -
   {\partial x^k\over \partial x'^j} {\partial v_{ i}\over\partial x^k} \Big)~.
   \label{bpatch}
\eea
Observe that the spacetime coordinates transform with the usual diffeomorphisms while the dual coordinates transform with a shift whose
parameter depends only on the spacetime coordinates. A modification of this proposal in the context of DFT was suggested in \cite{berman};
however, the transformations given in \cite{berman} reduce to the above after the strong section condition has been imposed.

Another proposal for the finite transformation of DFT was put forward  in \cite{rey}.  For this,  a closed 2-form was introduced $b$, $db=0$
which transforms as
\bea
b'_{ij}(x')= {\partial x^k\over \partial x'^i}\, {\partial x^l\over \partial x'^j} ((b_{kl}+\partial_k v_l-\partial_l v_k~)(x)),
\label{btrans}
\eea
while  ${\bf B}:=B-b$ is taken to transform tensorially:
\bea
{\bf B}'_{ij}(x')&=& {\partial x^k\over \partial x'^i}\, {\partial x^l\over \partial x'^j} \, {\bf B}_{kl}(x)
 ~.
 \label{bbtrans}
\eea
This implies that the $B$-field transforms as
\bea
B'_{ij}(x')&=& {\partial x^k\over \partial x'^i}\, {\partial x^l\over \partial x'^j} \, (B_{kl} +\partial_k v_l-\partial_l v_k)(x))
 ~,
 \label{bpatch2}
\eea
i.e. in the same way as $b$. In these equations $v$  depends only on the spacetime coordinates.
The doubled space coordinate transformations
are taken to be
\bea
x'{}^i=x'{}^i(x^j)~,~~~~\tilde x'_i=\tilde x_i+ v_i(x)~.
\label{dspacet2}
\eea

More recently a new proposal has been put forward \cite{hull}.  The doubled space coordinates transform as
\bea
x'{}^i=x'{}^i(x^j)~,~~~~\tilde x'_i=\tilde x_i~,
\label{dspacet3}
\eea
i.e. the spacetime coordinates transform with diffeomorphisms while the dual coordinates remain inert with respect to $B$-field gauge transformations.

The $B$-field transforms as

\bea
B'_{ij}(x')&=& {\partial x^k\over \partial x'^i}\, {\partial x^l\over \partial x'^j} \, \Big(B_{kl}(x)+\big(
 \partial_k v_{ l} -
 \partial_l v_{ k} \big)(x)\Big)
 ~,
   \label{bpatch3}
\eea
i.e. in the same way as in \cite{rey}.

  The reason that the dual coordinates $\tilde x$ do not transform under
the $B$-field gauge transformations is because the component $\tilde v_j$ of the generalised infinitesimal vector,
\bea
 V^M=\begin{pmatrix} v^i \\ \tilde v_j  \end{pmatrix}
\eea
that enters in the generalised Lie derivative acting on the fields, is identified as the parameter of an  infinitesimal gauge transformation of
 the $B$ field viewed as a gerbe connection. In other words $\tilde v_j$ is  viewed as (the parameter of) a gauge transformation rather than as a coordinate
 transformation.  Moreover the gerbe connection introduces a splitting in the short exact sequence
 \bea
 0\rightarrow T^*M\rightarrow E\rightarrow TM\rightarrow 0~,
 \eea
which describes the extension of $TM$ by $T^*M$. This allows  $E$ to be split as $E=TM\oplus T^*M$ and to thereby identify
the sections of $TM$ and $T^*M$ in $E$ which now transform as vectors and forms.  The calculation of how this can be done has been described
explicitly in \cite{hull} and amounts to going from $W$ generalised tensors to $\hat W$ ones in the notation of \cite{hull}.  This is related to the notion of the $B$-transform in generalised geometry \cite{hitchin2, gualtieri}.
As the dual coordinates of the doubled space $\tilde x$ do not transform, or just transform as 1-forms,  they are inert under $B$-field gauge transformations.
It has been argued in \cite{hull} that to describe DFT it is sufficient to consider the diffeomorphisms of the spacetime
together with the generalised geometry structure described above which includes a splitting of the exact sequence that determines the $B$ field.

\newsection{Patching}

Let us now turn to investigate the implications of patching doubled spaces with the
transformations proposed in the previous section on the topology and geometry of spacetime. Before we do this, let us describe a few
properties of the de Rham-\v Cech theory as applied to closed 3-forms $H$.  Let $\{U_\a\}_{\a\in I}$ a good cover, then on the open sets $U_\a$ and the n-fold
overlaps $U_{\a_0\dots \a_{n-1}}=U_{\a_0}\cap \dots \cap U_{\a_{n-1}}$, $n=2,3,4$, one has
\bea
&&H_\a=d B_\a~,~~~-B_\a+B_\b=d a^1_{\a\b}~,~~~a^1_{\b\g}-a^1_{\a\g}+a^1_{\a\b}=da^0_{\a\b\g}~,~~~
\cr
&&a^0_{\b\g\d}-a^0_{\a\g\d}+a^0_{\a\b\d}-a^0_{\a\b\g}=n_{\a\b\g\d}~,
\label{con1}
\eea
respectively, where $n_{\a\b\g\d}$ are constants.  The last condition arises from the requirement that on 4-fold overlaps
\bea
d(a^0_{\b\g\d}-a^0_{\a\g\d}+a^0_{\a\b\d}-a^0_{\a\b\g})=0~.
\label{con1a}
\eea
If $n_{\a\b\g\d}\in 2\pi\bZ$, then $H$ represents a class in $H^3(M, \bZ)$. The left-hand sides of all but the first of equations (\ref{con1}) involve the \v{C}ech differential $\delta$. It acts on form-valued fields defined on $p$-fold overlaps and takes them to forms on $(p+1)$-fold overlaps, e.g. $(\d B)_{\a\b}=-B_{\a} +B_{\b}; \ (\d a)_{\a\b\g}=a_{\a\b}+a_{\b\g}+a_{\g\a}$, and so on, and squares to zero, $\d^2=0$. \black

 We emphasise  that the 2-form gauge potential $B$ as well as the transition functions $a^1, a^0$ are not unique in the above decomposition. In fact
the decomposition is invariant under the local  ``gauge'' transformations
\bea
B_\a\rightarrow B_\a+d u^1_\a~,~~~a^1_{\a\b}\rightarrow a^1_{\a\b}- u^1_\a+u^1_\b+ d f^0_{\a\b}~,~~~a^0_{\a\b\g}\rightarrow a^0_{\a\b\g}+f^0_{\b\g}- f^0_{\a\g}+f^0_{\a\b}~,
\eea
where $u^1$ are 1-forms and $f^0$ are functions defined on the indicated overlaps.

\subsection{ $B$-dependent patching for dual coordinates}

If the coordinates for the doubled space, $\tilde x$, are taken to be one-forms patched together using the $B$-field transformations, i.e.

\bea
-\tilde x_\a+ \tilde x_\b \propto a^1_{\a\b}~,
\label{4.4}
\eea
as in \cite{lust} and \cite{berman} (where the notation $\zeta_{\a\b}$ was used for $a^1_{\a\b}$), then, as shown in \cite{gp1}, this implies that the $H$-flux is trivial. It follows from (\ref{4.4}) that $(\d a^1)_{\a\b\g}=0$ which can be solved by $a^1_{\a\b}=(\d u^1)_{\a\b}$ by the $\d$-Poincar\'e lemma. This in turn implies that $B_\a$ can be shifted by $(d u^1)_\a$ on each patch so that the new $B$-field  will be globally defined. So this construction of doubled spaces is not compatible with backgrounds with non-trivial $H$-flux in $H^3(M,\bZ)$. There are many examples of such backgrounds, for example those discussed in section 2.

Another patching proposal is that of \cite{rey} where  it is asserted  that the polarisation $b$, with $db=0$,  is defined on each patch $U_\a$ of a good cover  $\{U_\a\}_{\a\in I}$ and patches as\footnote{The notation $v_{\a\b}$ was used for $a^1_{\a\b}$ in \cite{rey}.}
\bea
b_\a=b_\b+da^1_{\a\b}~,
\label{bpatchx}
\eea
As $B$ transforms in the same way, the difference
${\bf B}=B-b$ transforms tensorially and one has $H=dB=d{\bf B}$, as $db=0$,  and so $H$ is exact.

One can reach the same conclusion  by  viewing the (\ref{bpatch2}) as a patching condition on a good cover as
\bea
(B_\alpha)_{ij}&=& {\partial x_\beta^k\over \partial x_\alpha^i}\, {\partial x_\beta^l\over \partial x_\alpha^j} \, (B_\b-b_\b)_{kl}+(b_\beta)_{ij}+(dv_{\a\b})_{ij}
   \label{bpatch22}
\eea
Since $b$ is closed, one can solve this locally as $b_\a=d u_\a$ and re-arrange the above equation using (\ref{bpatchx}) as
\bea
(B_\a-du_\a)_{ij}={\partial x_\beta^k\over \partial x_\alpha^i}\, {\partial x_\beta^l\over \partial x_\alpha^j} \, (B_\b-du_\b)_{kl}~.
\eea
However, as we have already mentioned the definition of $B$ is ambiguous up to a gauge transformation generated by $u$.  As a result $B$ can be chosen
to be a globally defined 2-form leading to an exact $H$.  This result is independent from the way that the dual coordinates transform
and so it is not affected by the gauge transformation introduced in \cite{park}.

An alternative reading of the proposal made in \cite{rey}, which is more tuned to the examples described later in that paper, is as follows.
One introduces
two different 2-form gauge potentials $B$ and ${\bf B}$ for the 3-form field strength $H$, but where now ${\bf B}$ is no longer necessarily tensorial.   If the transition functions
with respect to $B$ and  ${\bf B}$ are denoted by ${ a}^1$ and ${ a}^0$, and ${\bf a}^1$ and ${\bf a}^0$ in the \v Cech-de Rham decomposition, respectively,
we take the patching conditions of the dual coordinates $\tilde x$ to be those of the polarisation $b=B-{\bf B}$.  These are given by $a^1_{\a\b}-{\bf a}^1_{\a\b}:=\hat a^1_{\a\b}$.  So one can set
\bea
-\tilde x_\a+ \tilde x_\b=\hat a^1_{\a\b}~.
\label{4.8}
\eea
on each $U_{\a\b}$. This is similar to (\ref{4.4}) and implies that $\hat a^1_{\a\b}=(\d u^1)_{\a\b}$. So if we redefine $b$ by $b_\a\rightarrow b_\a-u^1_\a$ on each patch $b$ will be globally defined, while if we also redefine the new coordinates in a similar fashion, $\tilde x_\a\rightarrow \tilde x'_\a=\tilde x_a- u^1_\a$  the new coordinates will be inert under $b$ (or $B$)-field gauge transformations. This is similar to the first case discussed above, but now does not require that the flux of $H$ be trivial. So this interpretation leads to a patching condition which is equivalent to one which is  independent of the $B$-field patching.

\subsection{$B$-independent patching for dual coordinates}

Such a proposal is that described in   \cite{hull}. The patching conditions are just the diffeomorphisms of the spacetime and the patching conditions of the generalised geometry bundle $E$ together with a
choice of a splitting. The main point is that the patching conditions of the dual coordinates are
\bea
\tilde x_\a=\tilde x_\b
\eea
i.e. they remain inert. As the generalised geometry data are by construction globally defined, the patching of such a doubled space is consistent.

However, this proposal and in particular the assertion that $\tilde x_\a=\tilde x_\b$ is in conflict with the patching results that are a consequences of
the Buscher T-duality rules. As we have demonstrated with an explicit calculation
in section 2,     a T-dual circle can topologically twist over the spacetime. As this cannot happen to the DFT dual coordinates,
one can only conclude that according to this proposal DFT either does not incorporate the Buscher T-duality rules or the DFT dual coordinates $\tilde x$
should not be identified with the Buscher dual angular coordinates $\tilde \theta$.  If  the former is not considered desirable, then one must conclude that the
DFT dual coordinates $\tilde x$ is not the full story and additional coordinates must be introduced.  There has been such a suggestion before
in \cite{gp2} where the basis of generalised $\hat W$ generalised tensors has been identified and where it was shown how the generalised geometry emerges.
If this case, one might argue that the motivation for the introduction of the DFT dual coordinates in the first place is somewhat weakened, or that they have only
an auxiliary status.

To enforce the idea that a generalised geometry approach is not sufficient to describe the T-duality rules,
 observe that, although the generalised geometry bundle $E$ is twisted over the spacetime, as a space it is contractible to the spacetime
$M$. %\blue
In other words the spacetime is fixed and the bundle transformations, which one might wish to identify with T-duality transformations, cannot change the topology of the underlying space.
On the other hand, we have seen that T-duality changes the topology of spacetime, for example the sphere and the lens space have different cohomology groups, and moreover both spaces in the dual pair are smooth.
This does not mean that the T-duality transformation is necessarily smooth, but a smooth transformation of $E$ can never induce the T-dual geometry on the spacetime, i.e. only singular
gauge transformations of $E$ may be of interest as they may produce the desirable T-dual space.\black

The modified proposal of \cite{rey} discussed above also suffers a similar problem in that the patching condition (\ref{4.8}) does not reproduce the Buscher rules  and cannot accommodate dual angular coordinates.

\newsection{A new proposal}

\subsection{C-spaces and DFT coordinates}

Here we shall propose a scenario which illustrates the role of the various coordinates and how the Buscher T-dual spaces can be incorporated  using the C-space construction of \cite{gp2}. Given a good cover $\{U_\alpha\}$  of the spacetime $M$, one introduces new coordinates $y_\a^1$ on every open set $U_\a$ and angular coordinates $\theta_{\a\b}$ on every intersection
$U_{\a\b}$ and imposes the patching conditions
\bea
-y^1_\alpha+y^1_\beta+d\theta_{\alpha\beta}&=& a^1_{\alpha\beta}~,
\cr
\big (\theta_{\alpha\beta}+ \theta_{\beta\gamma}+ \theta_{\gamma\alpha}+  a^0_{\alpha\beta\gamma}\big)&=&0 ~~~\mathrm{mod}~ 2\pi \bZ~,
\label{c3space}
\eea
on $U_{\alpha\beta}$ and $U_{\alpha\beta\gamma}$.  Then consistency with  (\ref{con1}) at triple and fourfold overlaps implies
\bea
 n_{\alpha\beta\gamma\delta}=0 ~~~\mathrm{mod}~ \bZ~,
\eea
which is satisfied provided that ${1\over2\pi}H$ represents a class in $H^3(M, \bZ)$. The angular coordinates at double intersections are associated with the fibre
directions of the principal U(1) bundles that arise in the Hitchin-Chatterjee description of  gerbes \cite{chatterjee, hitchin}, explained in detail in \cite{murray}.

Common sector theories with $O(d,d)$ local gauge symmetry  can be described solely in terms of generalised geometry, i.e. without the introduction of additional coordinates.
Such theories can also be described in terms of C-spaces, as discussed in \cite{gp2}.
In this context  of C-spaces additional one-form coordinates can be introduced, as we have seen above, and it was  shown in \cite{gp2} that the first patching condition in (\ref{c3space}) can be used to introduce new one-form coordinates
\bea
\tilde y^1_\a=y^1_\a -\sum_\g \rho_\g (d\theta_{\a\g}- a^1_{\a\g})~,
\label{tildexy}
\eea
which are globally defined on the spacetime, i.e. $\tilde y^1_\a=\tilde y^1_\b$. Here $\{\rho_\a\}$ is a partition of unity subordinate to the good cover.
It seems reasonable on the grounds of their transformation properties to identify the $\tilde y^1_\a$ with the doubled coordinates
of \cite{hull} which also transform as one-forms, i.e. $\tilde x=\tilde y$,
after suppressing the degree and open set labels on $\tilde y$.
This incorporates the the DFT doubled coordinates into a  C-space description.

However, we have shown that the Buscher T-dual spaces cannot be  described in terms of the $(x, \tilde x)$ coordinates alone.   So the question that remains is where the Buscher T-dual spaces are hidden in this description. The C-space description contains in addition the angular coordinates $\theta$ which describe the  gerbe part of the space.  We shall argue that the Buscher T-dual spaces are hidden in the gerbe.

\subsection{Gerbes and Buscher rules}

Although in the construction of C-spaces a good cover has been used, for the definition of a  Hitchin-Chatterjee gerbe any open cover\footnote{Note, however, that for gerbes there is a notion of refinement  \cite{chatterjee}. As a result, any chosen open cover can be refined to a good open cover, so that any gerbe can be related to one defined on a good open cover.}
 suffices. We shall use this to adapt an open cover such that the Buscher T-duals can be described as subspaces of gerbes.

To illustrate how gerbes can be constructed,  consider the example of $S^3$ with $N$ units of $H$ flux. We have already seen that the T-dual space
of this is the lens space $L^3_N$ with one unit of flux.
 To  describe this gerbe on $S^3$ \cite{murray}, we can choose a stereographic cover of  two open sets $\{U_0, U_1\}$
on $S^3$  for which their intersection $U_0\cap U_1:=U_{01}$ is $I\times S^2$, $I$ an open interval, and the Mayer-Vietoris  description of  $H^3(S^3, \bZ)$ which uses  representatives localised
on $U_{01}$, see e.g. \cite{bott}.  Such representatives are constructed as follows. As $U_{01}$ is contractible to $S^2$, choose a representative $F_{01}$ of the class $Nu$ in $H^2(S^2\times I, \bZ)=H^2(S^2, \bZ)$ where $u$ is the generator
of  $H^2(S^2\times I, \bZ)$.  A representative of ${1\over 2\pi}[H]$ can be chosen as
\bea
\hat H_0=-d\rho_1\wedge F_{01}~,~~~\hat H_1=d\rho_0\wedge F_{01}~,
\label{5.4}
\eea
on $U_0$ and $U_1$, respectively, where $\{\rho_0, \rho_1\}$ is a partition of unity subordinate to the cover $\{U_0, U_1\}$.  Observe that at the intersection
\bea
-\hat H_0+\hat H_1=d (\rho_1+\rho_0)\wedge F_{01}= d 1\wedge F_{01}=0~,
\eea
and so $\hat H$ is globally defined on $S^3$.  Furthermore Stoke's theorem reveals that $[H]=[\hat H]$. As there are no more than double overlaps the rest of the compatibility conditions
for the gerbe are trivially satisfied. The gerbe\footnote{
  The gerbe in not a manifold.  From the perspective of $S^3$ it grows an extra dimension as one approaches the sphere at the equator. } associated to $S^3$ and $H$ is then the union of $S^3$
together with the principal $U(1)$ bundle on $U_{01}$ which has first Chern class $N u$.  Observe that the principal bundle over  $U_{01}$ when restricted on  $S^2\subset U_{01}\subset S^3$ is the Lens space $L^3_N$.  {\it It is significant that the lens space $L^3_N$ which is the T-dual to $S^3$  naturally appears in this gerbe construction.} Prompted by this, it is tempting
 to {\it identify the  T-dual angular coordinate $\tilde \theta$} with the {\it fibre coordinate of the lens space that appears in the gerbe construction}.  We shall provide a further
explanation for this below.

Suppose next  that the spacetime is a product $M=S^1\times Q$ and the 3-form flux $H=d\theta\wedge F$, where $F$ is a 2-form representing a class in  $ H^2(Q, \bZ)$ and where  we have suppressed
the pull back operation from $Q$ to $M$.
  Choose a cover on $S^1$ of two open sets $\{V_0, V_1\}$ then  $U_0=V_0\times Q$ and $U_1=V_1\times Q$
 are open and cover $M$, and their intersection $U_{01}= (V_0\cap V_1)\times Q$. As $F$ is defined on $M$  it  is also defined on $U_{01}$ and  we  denote its restriction to  $U_{01}$  by $F_{01}$.
Choose the gerbe principal $U(1)$ bundle  $P_{01}$  on $U_{01}$ to have Chern class represented by $F$.  Then a representative of the class of the 3-form flux $H$ on $M$ can be constructed as
in equation (\ref{5.4})
\bea
\hat H_0 =- d\rho_1\wedge F_{01}~,~~~\hat H_1=d\rho_0\wedge F_{01}~,
\eea
where
 now  $\{\rho_0, \rho_1\}$ is a partition of unity subordinate to the $\{V_0, V_1\}$ cover.
$\hat H$ is globally defined on $M$ and it is a representative of the 3-form flux associated to the gerbe. The T-dual space of $M$ is the bundle space of  $P_{01}$  which is
clearly a subspace of the  total space of  the gerbe.

 As a special case of the above take  $Q=T^2$.  In  this   case, $M=T^3$ and $H$ can be chosen as in section 2.4  in which  the T-dual pair of $T^3$ with flux was described.
In particular, we set
\bea
F=dp={N\over 2\pi} d\psi_2\wedge d\psi_3~.
\eea
In this case, the restriction of the principal $U(1)$ gerbe bundle $P_{01}$  on $Q=T^2\subset U_{01}\subset T^3$ is the T-dual space $\tilde T^3$ as described
by the Buscher T-duality rules. For a different treatment of this example, see \cite{hassler}.

 As a final example we take $M$ to be a circle bundle over $Q$ with 3-form flux $H$ that can be represented as $[H]= a w$, where $w\in H^2(Q, \bZ)$ and $a$ is the generator of $H^1(S^1, \bZ)$.
Take an open cover $\{W_\a\}$ on $Q$ which  trivialises the circle bundle $M$ over $Q$, i.e. $\pi^{-1}(W_\a)=\varphi^{-1}_\a(W_\a\times S^1)$, where $\pi:~ M\rightarrow Q$ is the
projection and $\varphi_\a: \pi^{-1}(W_\a)\rightarrow W_\a\times S^1$ is the trivilisation map, and write each $W_\a\times S^1$ as the union of the open set $W_\a\times V_0$ and $W_\a\times V_1$, where $\{V_0, V_1\}$ are the
two open sets that cover $S^1$ introduced  above. It is clear that $\{\varphi_\a^{-1}(W_\a\times V_r)\}$, $r=0,1$, is a cover for $M$. As the union of open sets is open
$U_0=\bigcup_\a \varphi_\a^{-1}(W_\a\times V_0)$ and $U_1=\bigcup_\a\varphi_\a^{-1}( W_\a\times V_1)$ are open and cover $M$. As in the case that $M$ was a product, we consider a representative $F$ of the class $w\in H^2(Q)$
and its pull back to $M$ with the projection map $\pi$.  Restricting $F$ to the intersection of $U_{01}$ and denoting it by $F_{01}$, we can construct a representative $\hat H$
of the $H$ flux as in (\ref{5.4}),
where again $\{\rho_0, \rho_1\}$ is a partition of unity subordinate to the cover $\{U_0, U_1\}$. $\hat H$ is globally defined and represents $[H]= a w$ as the derivatives
of the partition functions at the intersection of open sets on the circle represent the generator of $H^1(S^1, \bZ)$. It is clear that the gerbe is the union of spacetime with a circle bundle defined on the open set $U_{01}$
 of $M$  which is  the restriction of the pull-back of a circle bundle over the base space $Q$ with Chern class $w$. The circle bundle over $Q$ is the T-dual
space derived from the Buscher rules. If $S^3$ is viewed as a circle fibration over $S^2$ and $H$ represents $N$ units of flux, the above gerbe construction
will also lead to the identification of the T-dual space as $L_N^3$.  It is clear that the gerbes in all the above examples have simple descriptions because
the spacetimes have been covered by only two open sets.

\subsection{Summary of the proposal}

The above results provide evidence to suggest that the double coordinates $\tilde x$ of DFT that transform like 1-forms \cite{hull} should be identified
with the $\tilde y$ coordinates that occur in C-spaces, eqn (\ref{tildexy}).  DFT can be formulated with only these coordinates and will exhibit local
$O(d,d)$ symmetry as such a description  accommodates generalised geometry both from the double spaces point of view and that of C-spaces.  However, such a formulation will not describe the T-dual spaces of the spacetime.
This is regardless of the choice of solution to the strong section condition that one makes on the doubled space.

Our results have also established that the T-dual space of a spacetime with $H$ flux  can be identified as a  subspace of a  gerbe which is part of the C-space. This has been done explicitly for
 the T-dual space derived after performing T-duality along the fibre direction of  a spacetime which is a circle fibre bundle.  This identification requires the presence of additional coordinates from those of doubled space
which are the fibre coordinates of the principal $U(1)$ bundles
that lie on double  intersections of an open cover of  the spacetime and  are required in the description of the gerbe.

We have given two gerbe descriptions of  the T-dual space of  $S^3$  with $N$ units of $H$ flux example.  The first description was in terms of
a stereographic cover and the other in terms of a cover adapted to the fibration over $S^2$.  In both cases, the T-dual space has been identified as the
lens space $L_N^3$.  In hindsight this may have been expected.  The T-dual space should be independent from a large enough selection of covers on the spacetime
that are used to describe the gerbe.
This can be seen as the requirement for the construction of gerbes and that of T-dual spaces to be covariant.   In turn one can view this as  a covariant description of the Buscher T-duality rules.

\newsection{Conclusions}

We have made a proposal based on C-spaces and  the Hitchin-Chatterjee description of a gerbe where both the local (bundle) $O(d,d)$ symmetry and the T-dual spaces of a spacetime can be described
in a globally consistent way. In particular, we have demonstrated that the doubled space of a DFT as described in \cite{hull} can be included into a C-space and  the $O(d,d)$
symmetry arises as part of the generalised geometry structure on C-spaces. Furthermore, we have demonstrated that the T-dual spaces of a spacetime that are  constructed
using Buscher rules can be identified  as subspaces of the gerbe which is included in  C-spaces but not in the doubled spaces. In this identification,
the T-dual angular coordinate of a spacetime which is a circle fibration with  T-duality operation taken along the fibre circle is identified with the gerbe angular coordinate  which is
the fibre coordinate of a principal $U(1)$ bundle defined on an intersection of two open  sets of the spacetime.

Our analysis has indicated   that it is not possible to formulate a theory which exhibits both local $O(d,d)$ symmetry and at the same time
has a description of all the T-dual spaces  of a spacetime based only on  doubled spaces. Using the available globally consistent definitions of doubled spaces, we have demonstrated that  these cannot provide an explanation for the property of  the T-dual circles to topologically twist over the spacetime.
This topological twisting has been established in several examples and it is a consequence of the Buscher rules. In other words, the T-dual spaces
cannot arise in DFT as different solutions to the (strong) section condition on doubled spaces.

The inclusion of gerbes in a consistent definition of a theory which exhibits  local $O(d,d)$ symmetry and which describes the T-dual spaces of a spacetime
requires the presence of additional angular coordinates, the gerbe coordinates.  Such spaces are not manifolds
and in particular they do not have a fixed dimension.  Nevertheless they contain all the necessary ingredients for the definition of the theory
including  the ability to perform differential geometry computations related to $O(d,d)$ symmetry and the topological properties required for the descrption
of the T-dual spaces.

The gerbe description of T-dual spaces of a spacetime has some additional consequences.  First notice that the Buscher rules are not covariant.
Their formulation  involves several gauge choices and their construction is essentially local on the spacetime. Moreover, they depend on
the spacetime admitting an isometry.  On the other hand gerbes can be defined on any smooth manifold  with a
closed 3-form flux $H$ without further additional assumptions.  Therefore the gerbe description can be seen as a covariantisation  of
the T-duality rules.  Furthermore the gerbe description opens the possibility  that it might be possible to investigate the T-duals of a spacetime that does not admit
isometries.  In this case, however, it may not be possible to  identify the subspaces of the gerbe which can be characterised as  T-dual spaces as we have done in the case
of spacetimes with isometries. Even if the T-dual spaces can be identified, it is likely that they will not be manifolds.

\vskip 1cm

\noindent{\bf Acknowledgements} \vskip 0.1cm
GP is partially supported by the  STFC rolling grant ST/J002798/1.

\setcounter{section}{0}\setcounter{equation}{0}

% \appendix{Modifying transition functions}

\setcounter{section}{0}\setcounter{equation}{0}

\end{document}